\documentclass[useAMS,usenatbib]{elsart}

\usepackage{graphicx}
\usepackage{journals_abrv}

\newcommand{\R}{\mathcal{R}}

\def\apgt{\ {\raise-.5ex\hbox{$\buildrel>\over\sim$}}\ }

\newcommand{\one}{\#1}
\newcommand{\two}{\#2}
\newcommand{\three}{\#3}

\journal{Communications in Nonlinear Science and Numerical Simulation}
\begin{document}
\begin{frontmatter}
\title{Detecting Irregular Orbits in Gravitational \emph{N}-body Simulations}

\author[API,SCS]{J. F. Gemmeke}
\author[API,SCS]{S.F. Portegies Zwart}
\author[API]{C. J. H. Kruip}
\address[API]{Astronomical Institute 'Anton Pannekoek', University of Amsterdam, Amsterdam, the Netherlands}
\address[SCS]{Section Computational Science, University of Amsterdam, Amsterdam, the Netherlands}



\begin{abstract}

We present a qualitative diagnostic based on the continuous wavelet
transform, for the detection of irregular behaviour in time series of
particle simulations.  We apply the method to three qualitatively
different gravitational 3-body encounters. The intrinsic irregular
behaviour of these encounters is well reproduced by the presented
method, and we show that the method accurately identifies the
irregular regime in these encounters.  We also provide an
instantaneous quantification for the degree of irregularity in these
simulations.  Furthermore we demonstrate how the method can be used to
analyze larger systems by applying it to simulations with
100-particles.  It turns out that the number of stars on irregular
orbits is systematically larger for clusters in which all stars have
the same mass compared to a multi-mass system.  The proposed method
provides a quick and sufficiently accurate diagnostic for identifying
stars on irregular orbits in large scale $N$-body simulations.

\end{abstract}


\end{frontmatter}

\section{Introduction}\label{sec:intro}

Gravitational $N$-body systems are intrinsically chaotic, at least for
$N>2$ \cite{2003Ap&SS.283..347C}. For $N=2$ Newtonian
systems are regular,
but post Newtonian
dynamics\cite{1915SPAW.......778E,1915SPAW.......799E,1916AnP....49..769E}
can already reveal chaotic behaviour
\cite{1978MitAG..43Q.121B,2003PhRvL..90m4101B} by $N=2$.  Chaoticity
in a self gravitating systems with $N>2$ can reveal itself on a very
short time scales, of the order of orbital periods.  On the other
hand, in some systems like the planetary system around the Sun,
chaoticity only reveals itself on a time scale of billions of orbits
\cite{1860MNRAS..21...60D,1860MNRAS..20..240D}. In a stellar cluster
which does not contain a dominant central mass, orbits can be chaotic
on a much smaller time scale. In such an environment seemingly regular
orbits can suddenly become highly irregular, to return later to
regular again \cite{1962AJ.....67..591K,1979IAUS...81..231K}.

In this study we focus on the characterisation of orbits which show
irregular behaviour on a short (dynamical) time scale, and not on a
long (relaxation) time scale.  Star clusters, with contain between
about 100 and {\large O($10^7$)} stars, are governed by microscopic few
body interactions \cite{1996magr.meet..167K}.  Analysis of such
systems is often hampered by this internal microscopic physics and a
qualitative indicator of chaotic behaviour could enormously assist in
the understanding of large scale gravitational $N$-body simulations.
In particular, since it is thought that stars on irregular orbits in
self gravitating $N$-body systems have an important effect of the bulk
properties of such systems \cite{2002SSRv..102..115M}.

We present a transparent diagnostic to qualify chaotic behaviour of
individual trajectories in $N$-body systems. In addition, the method
has some quantitative qualities. The application of this method ranges
from 3-body interactions to simulations of entire star clusters and
galaxies.  For clarity, we define irregular orbits as orbits which are
deterministic though sensitive to the initial conditions and which
cannot be described as a sum of periodic motions.

The motion of a star in an \emph{N}-body simulation can change from
regular to irregular (and vice versa) due to gravitational
interactions with other stars.  Irregularity of an orbit then is a
local quantity, and this forces us to deviate from the conventional
methods based on Lyapunov \cite{Lyapunov:1901} numbers such as
explained in \cite{1997CeMDA..67...41F,Brasser:2004,Sandor:2004}.  In
addition, the calculation of Lyapunov exponents is costly, may require
reruns and are therefore less suited for a direct diagnostic, whereas
we are predominantly interested in a diagnostic that can be evaluated
at runtime. The Lyapunov indicator, however, is well suited for
discriminating between ordered and weak chaotic motion, like planetary
systems, whereas we are in particular interested in very chaotic
systems \cite{1997CeMDA..67...41F}. Others methods for detecting chaos
in Hamiltonian systems, such as SALI
\cite{Kalapotharakos:2004,Skokos:2004}, Fourier Transform
\cite{Aguilar:1998,Laskar:1998,Meritt:1998}, Poincar\'e Section
\cite{Poincare:1892}, the zero-one method \cite{Gottwald:2002} and the
geometric indicator \cite{Cipriani:2002} are also less suited for
analysing gravitational $N$-body simulations during run time, since
they also are not practical in providing an instantaneous
quantification of the chaoticity of the system we are interested in
here.

In \S\,\ref{sec:method} we discuss the basics of the Continuous
Wavelet Transform (CWT), which we present in a package called CWaT. We
then proceed by applying this method to several \emph{N}-body
simulations in \S\,\ref{sec:application}, to conclude in
\S\,\ref{Sect:Conclusions}.

\section{Method for detecting chaos}\label{sec:method}

\subsection{The continuous wavelet transform}\label{subsec:method:CWaT}
The continuous wavelet transform \cite{Goupillaud:1984} (CWT) gives
the time-frequency representation of a time series $f(t)$ by fitting a
wavelet $\Psi$ to it at subsequent points in time $t$. We have used
the Morlet-Grossmann wavelet \cite{Grossmann:1984}, which is described
by

\begin{equation}\label{eq:Morlet}
        \Psi(t)=e^{i\omega_{0}t}e^{{-t^2}/{2\sigma^2}}.
\end{equation}

\noindent Here $\sigma$ is a measure of the spread in time and
$\omega_{0}$ the base frequency of the wavelet. The wavelet is a
periodic sinusoidal signal enveloped by a Gaussian, from which we can
construct a family of wavelets using a scaling factor $\alpha$ and a
time shift $\beta$
\begin{equation}\label{eq:family}
        \Psi_{(\alpha,\beta)}(t) = \frac{1}{\alpha}\Psi
                                   \left(\frac{t-\beta}{\alpha}\right), 
                                   \qquad \quad \beta \in \R, 
				   \qquad \quad \alpha > 0.
\end{equation}
We define the CWT as

\begin{equation}\label{eq:CWaT}
        T(\alpha,\beta) = \int\limits_{ - \infty }^\infty  
	                  { f(t)\overline{\Psi}_{(\alpha,\beta)}(t) dt }.
\end{equation}
Here $\overline{\Psi}$ is the complex conjugate of $\Psi$ and $f(t)$ a
time series.  Fitting the wavelet to $f(t)$ at discrete moments in
time ($\beta \in [0,t]$) yields $T(\alpha,t)$. We express
$T(\alpha,t)$ in terms of frequency $\omega$ rather than scale
$\alpha$ (since we are more accustomed to the former) using the
relation

\begin{equation}
\label{eq:scalefrequency}
        \omega=\frac{\omega_{0}}{\alpha}.
\end{equation}
This leads to a time-frequency representation $T(\omega,t)$. For an
extensive discussion see \cite{Mallat:1998}.

\subsection{Analysing the CWT}\label{subsec:method:ridges}

The main features of the time series can be described by its
instantaneous frequencies $\omega(t)$.  These can be extracted from
$T(\omega,t)$ by selecting its local maxima.  We only consider all
local maxima larger than a threshold $\tau = 0.95$ of the global
maximum.  In this we take a different approach than
\cite{Nesvorny:1996,Arevalo:2004} who explicitly consider the global
maximum. Using a lower threshold enables us to filter most of the noise
from the CWT without losing the relevant information on which the
$\chi$ can be determined. We also tried $\tau = 0.93$ and $\tau =
0.97$, which gave very similar results in the resulting value of
$\chi$.

We find the extrema by simulated annealing
\cite{1983Sci...220..671K,Cerny:1985}, following \cite{Carmona:1998}.
Connecting the local maxima results in a number of curves in the
time-frequency domain, which we call {\em ridges}.  The time series is
subsequently analysed by studying the ridges, enabling a qualitative
analysis of the behaviour of the time series. This is considerably
more practical than using the CWT directly, an example is presented in
fig.\,~\ref{fig:non_res_pres}.

A periodic time series (i.e. with a constant frequency) results in a
uninterrupted horizontal ridge indicating the fundamental frequency
$P$ and possibly overtones spaced at periods $P/n$ where $n$ is
integer.  A quasi-periodic time-series can be described by a sum of
periodic time series, each with its own ridge at a specific frequency.
Such a time series is then represented by multiple horizontal
ridges. Irregular time series on the other hand lead to curved ridges
of limited length \cite{Chandre:2003}.

\subsection{A quantitative measure of chaos}\label{subsec:method:implement}

To bring the chaos detecting qualities of the method one step further
we quantify the number of ridges and their curvature in the
irregularity indicator $\chi$. We define $\chi$ at any moment in time
by:
\begin{equation}\label{eq:chaosmeasure}
       \chi(t) = \frac{1}{n(t)}\sum\limits_{i=1}^{n(t)} 
	         \left|s_{i}\right|.
\label{eq:chaos}
\end{equation}
Here $n(t)$ is the number of ridges at time $t$ and $s_{i}$ a measure
for the slope of ridge $i$.  Using a least squares fit we calculate
the slope of the ridge.  Because $\chi$ would be dominated by large
values of the slope we use an empirical cutoff.  The slope is
normalised to unity to assure that $s_{i} \in [0,1]$.

The ridges of periodic and quasi-periodic time series are represented
by horizontal ridges, they imply $\chi(t)\simeq 0$ (see
\S\,\ref{subsec:method:ridges}).  We note however that the proposed
indicator (see Eq.\, \ref{eq:chaosmeasure}), cannot give a yes/no
definition about the occurrence of local exponential instability, but
it is capable of giving a measure of the degree of instability.

\section{Application}\label{sec:application}

To demonstrate the effectiveness of CWaT we apply it to
three gravitational 3-body interactions. Such interactions are
irregular \cite{1876QB43.D34.......}, with some regular
characteristics (see \cite{1993PhDT.........2B,1994CeMDA..58....1A}
for an interesting earlier analysis of this sort).

In each of the three following examples we inject a single object
(star) in a system consisting of two objects (stars) which are bound
in a binary.  A time series is constructed by projecting the orbit of
each of the stars on three orthogonal axes (see also
\cite{Aguilar:1998}).  The stellar orbits are computed using the {\tt
starlab} software environment \cite{SPZ:2001,2004Natur.428..724P}
which is publicly available from \\ {\tt
http://www.ids.ias.edu/\~{}starlab}.  This $N$-body integration
package calculates the orbits of the stars with individual time steps
\cite{Aarseth:1985} using a fourth-order Hermite predictor-corrector
scheme \cite{Makino:1992}. A regular time series is constructed by
Hermite interpolation at evenly spaced time intervals.  The time step
adopted for CWT is $1/1024^{\rm th}$ of the total integration
time. Further analysis is carried out with the S-WAVE software package
\cite{Carmona:1998}, to ultimately calculate the CWT and the ridges of
the time series.

In the following two \S\S\, we discuss the chaotic behaviour in one
non-resonant preserving encounter and one resonant encounter.  In the
first encounter the incoming star escapes after a close encounter with
a binary.  In the following example, in
\S\,\ref{subsec:discuss:dem_res_pres} (see \cite{Hut:1983} for more
examples, and even some very peculiar orbits in
\cite{2003gmbp.book.....H}), the incoming star has multiple close
interactions with the binary, we call such interactions `resonant'.
The example shown is a democratic resonance which is characterised by
all three stars interacting strongly with multiple close passages;
each star generally interacts with comparable frequency with any of
the other stars.

We use the following convention: star \one\, and \two\, form the
initial binary which is encountered by star \three. The time series
give the position of the star with respect to the centre of mass of
the 3-body system.  The units are given in standard \emph{N}-body
units \cite{Heggie:1986}.

\subsection{Irregular motion in a non-resonant gravitational interaction}
\label{subsec:discuss:non_res_pres}

In the first example we select an interaction between a binary,
consisting of star \one\, and \two\, to interact with a single
incoming star, \three.  After interacting with the binary, star
\three\, escapes without disrupting the binary. The incomer, however,
may change the binding energy of the binary.  Precise initial
conditions are presented in Tab.\,\ref{tab:non_res_pres}.

\begin{figure}
        \includegraphics[scale=0.2]{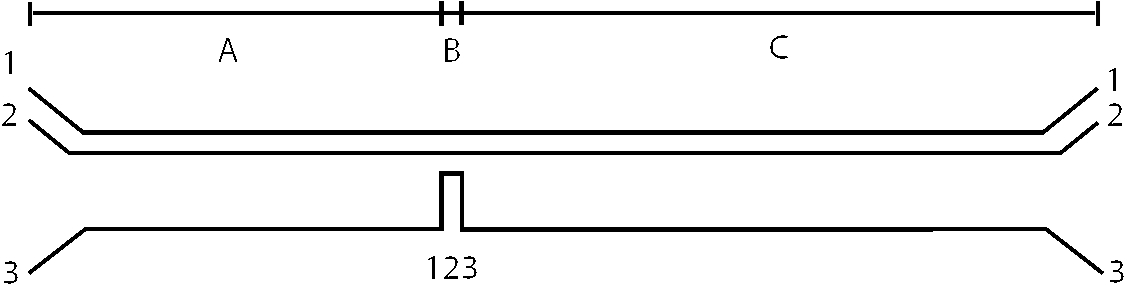}
        \caption{\label{fig:non_res_pres_knot} Schematic
        representation of a non-resonant three body interaction.  The
        interacting stars on the left are indicated by 1, 2 and 3,
        each of which progresses in time along the solid curve to the
        right. Initially star 1 and 2 are close together, as is also
        indicated in the representation. A close encounter with star 3
        sometime half way the encounter is marked by the three lines
        coming close together.  The label ``$123$'' identifies the
        moment of closest approach between the three stars. To the top
        of the schematic representation we plotted a bar with the
        letters `A' through `C', which correspond to various regions of
        different behaviour. A more quantitative analysis is presented
        in fig.\,\ref{fig:non_res_pres}
}
\end{figure}

In fig.\,.\,\ref{fig:non_res_pres_knot} we present a schematic diagram
of the interaction. In this example the binary has a close encounter
with the incomer at $t \simeq 150$. After the encounter the incoming
star is deflected along a hyperbolic path.

\begin{figure}
        \includegraphics[scale=0.57,angle=-90]{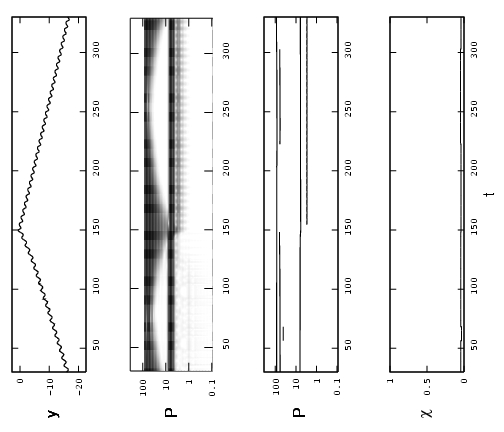}
        \caption{\label{fig:non_res_pres} Analysis of the non-resonant
        3-body interaction. The schematic interaction is presented in
        fig.\,~\ref{fig:non_res_pres_knot} and the initial conditions
        are provided in tab.\,1.  The top panel gives the time series
        $y(t)$ of star \two. The sinusoidal line to the bottom right
        shows a projection of the orbital motion of star \two.  The
        second panel contains the CWaT for this time series, followed
        by the ridge extraction in the third panel from above.  The
        bottom panel shows the $\chi$ according to
        equation~\ref{eq:chaos}.}
\end{figure}

The analysis using CWaT of one of the nine (three stars and three
axes) time series is given in fig.\,\ref{fig:non_res_pres}. The top
panel shows the time series of the position of star \two\, projected
on the $y$-axis in \emph{N}-body spacial coordinates versus
\emph{N}-body time units.  In the first $150$ \emph{N}-body time
units, $t=[0,150]$, the binary moves upward (in $y$) with respect to
the centre of mass because the incomer (star \three) is approaching it
from above. The projection of the circular motion results in the
sinusoidal behaviour of the time series.  The motion of the binary with
respect to the centre of mass is downwards after the encounter with
the incomer.

The second panel of fig.\,~\ref{fig:non_res_pres} shows
$T(\omega,t)$. Its most prominent feature is two arch-like shapes in
the period region 10 to 100 spanning $t = [30,150]$ and $t =
[150,330]$.  These arches are the sum of one arch spanning $t=
[30,330]$ due to boundary effects \cite{Chandre:2005} and a triangular
feature centred at $t \simeq 150$ due to the centre of mass motion of
the binary.  These regions of high intensity result in ridges with
periods $P \simeq 86$ and $P \simeq 61$ in the third panel.

The ridge at $P \simeq 6.3$ in the third panel indicates the period of
the binary system, which after $t\simeq 150$ has reduced to $P \simeq
6.0$.  This ridge is joined by another on the interval $t=[150,330]$
at a shorter period of $P\simeq3$. The lower period ridge is an
overtone caused by an increase of the eccentricity of the binary from
$e=0$ to $e\simeq0.38$.

As we discussed in Section \ref{subsec:method:implement}, the bottom
panel of fig.\,~\ref{fig:non_res_pres}, shows that $\chi \simeq 0$
during the entire interaction in accordance with the periodic
behaviour of the system. The short fly-by interaction has not let to a
qualitative indication of irregular behaviour in the system.

A similar analysis can be performed for any star and for each
projection. The result will be quite similar, supporting the argument
that the presented method is independent of the selection of a
fundamental plane.

\DeclareGraphicsRule{.jpeg}{eps}{.jpeg.bb}{`jpeg2ps -h #1}

\DeclareGraphicsRule{.jpeg}{eps}{.jpeg.bb}{`jpeg2ps -h #1}

\subsection{Irregular motion in a democratic resonant interaction}
	\label{subsec:discuss:dem_res_pres}

\begin{figure}
        \includegraphics[scale=0.2]{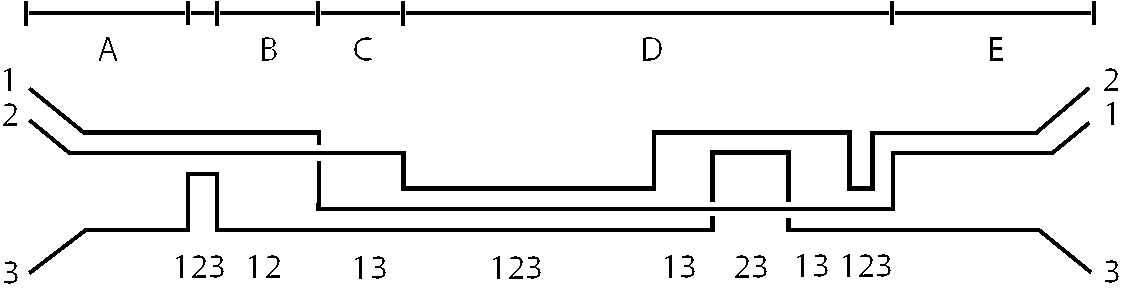}
        \caption{\label{fig:dem_res_pres_knot}
Schematic representation for the democratic resonant 3-body
interaction. The analysis of this interaction is given in
fig.\,~\ref{fig:dem_res_pres}.}
\end{figure}

\begin{figure}
        \includegraphics[scale=0.57,angle=-90]{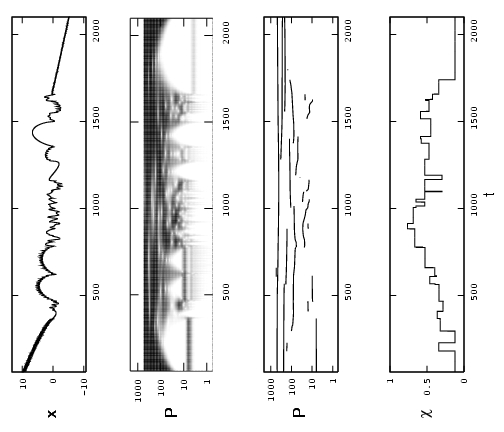}
        \caption{\label{fig:dem_res_pres}
Analysis of the democratic resonant 3-body interaction from
fig.\,\ref{fig:dem_res_pres_knot}.  The analysis is based on the
time-series $x(t)$ of star \one.}
\end{figure}

In a democratic resonance encounter the incoming star interacts
closely with both components of the initial binary, until one of the
stars eventually escapes.  The initial conditions used for this
example are presented in Tab.\,2 and a schematic representation of the
interaction is shown in fig.\,~\ref{fig:dem_res_pres_knot}, and the
analyse using CWaT is presented in fig.\,~\ref{fig:dem_res_pres}.

This interaction is considerably more complicated that the previous
two examples.  In fig.\,\ref{fig:dem_res_pres_knot} and
\ref{fig:dem_res_pres} we see how the incomer, star \three, is
captured by the binary (consisting of star \one\, and \two) at $t
\simeq 400$ and interacts several times strongly with the other two
stars before being ejected again at $t \simeq 1700$.  The initial
binary has a period $P\simeq 6.3$, but after $t \simeq 400$ the CWT
becomes very irregular.  The most irregular regime is noticeable in the
analysis for $\chi$ in the bottom panel of
fig.\,\ref{fig:dem_res_pres} at $t = [450,1700]$, with a peak near $t
= [750,1000]$, where values of $\chi \apgt 0.6$ are reached.

\DeclareGraphicsRule{.jpeg}{eps}{.jpeg.bb}{`jpeg2ps -h #1}

\DeclareGraphicsRule{.jpeg}{eps}{.jpeg.bb}{`jpeg2ps -h #1}

The strong democratic resonance interaction in this \S\, provides an
excellent example to compare the CWaT method with Poincar\'e sections,
which we will do in the subsequent \S.

\subsection{Poincar\'e section of a democratic resonant interaction}
\label{subsec:comparison}

Poincar\'e sections have, like the CWaT diagnostic, the advantage that
a qualitative impression of the irregularity in time series can be
obtained instantaneously, id est; at run time.  The main disadvantage
of Poincar\'e sections is the required choice of a fundamental plane.
This hinders analysis of complicated orbits in larger $N$-body
systems, as there is no simple automated procedure to find a
fundamental plane through which the Poincar\'e section can be taken.
This is in particular a problem in systems with hierarchical
substructure, such as the motion of moons around planets which again
move around a star.  The CWaT method does not suffer from this
complication and, in addition, provides an unbiased indicator for the
degree of irregularity.  We here compare the CWaT method with
Poincar\'e sections to demonstrate that global behaviour is reproduced
in both methods.

\begin{figure}
        \includegraphics[scale=0.55,angle=-90]{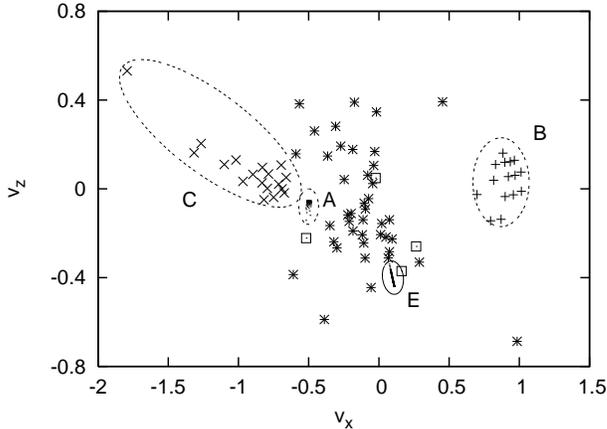}
        \caption{\label{fig:drp_poinc} 
	Poincar\'e section of the velocity components of the
        democratic resonant 3-body interaction shown in
        fig.\,~\ref{fig:dem_res_pres_knot} and further analysed in
        fig.\,\ref{fig:dem_res_pres}. Here we show $v_{x}$ and $v_{z}$
        for star \one\ at each moment that $v_{y}$ becomes
        negative. The letters `A', `B', `C' and `E' correspond to the
        regions indicated in
        fig.\,~\ref{fig:dem_res_pres_knot}. Scattered asterisks
        between region `B' and `C' belong to time interval indicated
        with `D'. The little squares indicate the short intermittent
        period between `A' and `B'.
}
\end{figure}

Poincar\'e sections are computed using the velocity components of a
star along a fundamental plane and record the moments a star passes
through that plane. A stable binary in this representation gives a
linear relation between the velocity components at each intersection.
For the democratic resonant interaction of
\S\,\ref{subsec:discuss:dem_res_pres} we use the velocity components
of star \one\, in the $v_{x}v_{z}$-plane at the moments $v_{y}$
becomes negative.  The centre of mass of the 3-body system remains in
the origin throughout the simulation. The results of this are presented
in fig\,~\ref{fig:drp_poinc}.

In fig\,~\ref{fig:drp_poinc} we indicate the five epochs by the
letters 'A' through 'E'. The same designation is used in
fig.\,\ref{fig:dem_res_pres_knot}.  The points is regions 'A' and 'E'
are clustered together, indicating that the orbit of the star before
('A') and after ('E') is well behaved.  During the resonant encounter
('B', 'C' and 'D') each of these resonant epochs can be grouped with
hindsight from the CWaT method. In fig.\,\ref{fig:drp_poinc} we use
different symbols to identify the various resonant epochs.

\subsection{Irregular motion in a 100-body system}
            \label{subsec:100-particles}

After having confirmed that the CWaT method enables a reliable
detection of irregular orbits in small N-body systems it is time to
focus on the larger systems. Here we will apply CWaT to self
gravitating $N$-body systems with 100 particles.  In particular we
report here the result for several of such simulations.  The initial
realizations of the particle clusters ware generated as follows.  We
make a distinction between two basic models, model EqM and model
S. Model EqM was constructed by distributing 100 objects of the same
mass in a Plummer \cite{Plummer:1911} Spherical density distribution.
The velocity of all point masses are subsequently scaled such that the
entire system is in virial equilibrium, and to $N$-body units
\cite{Heggie:1986}.  We run the system for 100 $N$-body time units,
which corresponds to about 35 dynamical time units.

\begin{figure}
		\includegraphics[scale=0.4,angle=-90]{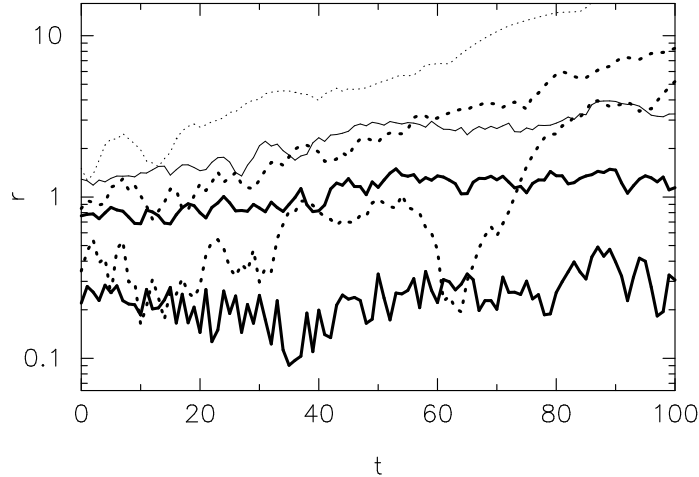}  
		\caption[]{ 
   Evolution of the radii which contains 5\% (bottom thick solid
   curves), 50\% (middle solid curve) and 75\% (thin solid curve)
   of the mass for simulation model EqM (all masses are equal).  For
   model S, with a Salpeter initial mass distribution for the
   individual particles, the same data is presented as the thick
   dotted curve (5\% Lagrangian radius) and the two thin dotted curves
   representing the 50\% and the 75\% Lagrangian radii.
\label{Fig:Lagrangian_radii}
} 
\end{figure}

In fig.\,\ref{Fig:Lagrangian_radii} we present the evolution of
several fundamental radii for simulation model EqM which contain 5\%,
50\% and 75\% of the entire mass of the system. Although it is not our
intention to discuss the dynamical evolution of this system it is
worth mentioning that the core experiences a gravothermal runaway
\cite{1983MNRAS.204P..19S} at about $t \simeq 35$, noticeable as the
deepest point reached by the lower solid curve (5\% Lagrangian
radius), whereas the outer parts of the cluster continues to expand,
mainly due to binary heating.

\begin{figure}
\includegraphics[scale=0.4,angle=-90]{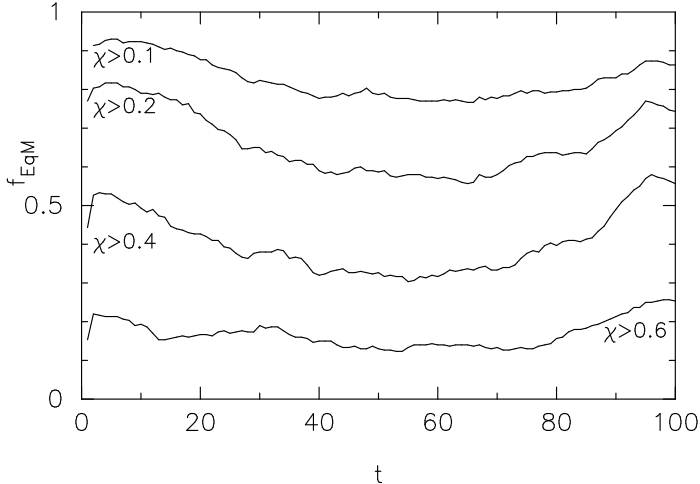} 
\caption[]{ 
  The evolution of $f_{\rm EqM}$ for simulation model EqM, the
  fraction of stars with $\chi > 0.1$, 0.2, 0.4 and 0.6 (from top to
  bottom).  
\label{Fig:Eqm_Chi}
} 
\end{figure}

A further analysis of simulation EqM using CWaT, is presented in
fig.\,\ref{Fig:Eqm_Chi}. For each star we compute the CWT for the
duration of the simulations. In fig.\,\ref{Fig:Eqm_Chi} we then show
the fraction of stars ($f_{\rm EqM}$) which has a $\chi > 0.1$ (top
curve), 0.2, 0.4 and 0.6 (bottom curve). The various curves are rather
smooth and flat.  Some caution is well taken here, though, as near the
edges (within the first and last $\sim 20$ time units) the CWaT
becomes unreliable.  The range between $t=20$ and $t=80$, where the
method is accurate, we notice that the fraction of stars with any
degree of chaoticity is roughly constant. It is interesting to note
that $\apgt 20$\% of the stars have $\chi > 0.6$ and that 30--40\% of
the stars have a $\chi < 0.2$. In
\S\,\ref{subsec:discuss:dem_res_pres}, by the way, we demonstrated
that a value of $\chi \apgt 0.6$ is a strong indication that the
underlying orbits change on a very short time scale.  The core
collapse around $t \simeq 35$ is not clearly noticeable. Core collapse
has apparently no direct influence in the number of irregular orbits
in the equal-mass $N$-body system.

We run a second simulation with the same number of particles,
distributed initially using the same 3 dimensional density profile,
in which we assign randomly masses to each of the stars from a
power-law distribution. We opted for a \cite{Salpeter:1955} mass
function which has the form of $N(m) \propto m^{-2.35}$.  Here $N(m)$
is the number of stars formed per unit mass range. We apply this mass
function over a range of two orders of magnitude, i.e, the most
massive star is a hundred times more massive than the least massive
stars. We call this model S.

We perform several simulations using this mass function but with new
initial realizations of the system.  The run-to-run variations can be
quite large, mainly due to the small number of stars.  A consistent
trend, however, is evident.

The dotted curves in fig.\,\ref{Fig:Lagrangian_radii} present the
evolution of the fundamental radii containing 5\%, 50\% and 75\% of
the mass of simulation model S.  The evolution of the structure of
model S is quite different than for the simulation with equal masses
(model EqM). In the former case core collapse occurs much earlier (at
about $t = 14$), which is consistent with simulations of
\cite{2000IJoMP...15..4871P}. However, model S also shows a second
core collapse near $t \sim 63$.  Noticeable is the much more
dramatic expansion of the outer parts of the cluster, which is driven
by the heating of the cluster core by close bound pairs which form
during the core collapse (exempli gratia see
\cite{2003gmbp.book.....H,Aarseth2003}).

\begin{figure} 
\includegraphics[scale=0.4,angle=-90]{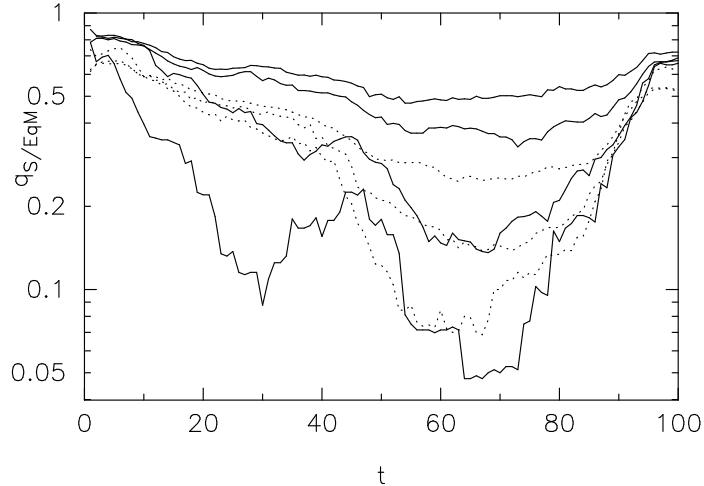} 
\caption[]{ 
  The evolution of the ratio of $q \equiv f_{\rm S}/f_{\rm EqM}$. The
  solid and dotted curves are for two different realizations of the
  initial conditions of the $N$-body system with initial parameters
  according to model S.  From top to bottom the curves are computed by
  adopting $\chi > 0.1$ (top solid and dotted curves), 0.2, 0.4 and
  $\chi > 0.6$ (bottom sold curve). To prevent clutter the latter
  value for $\chi$ is presented only for one of the simulations.
\label{Fig:Chi_ratio}
} 
\end{figure}

We compute the CWT for each star in model S, for which we show the
results of two simulations. For a better comparison we present the
ratio $q_{\rm S/EqM} \equiv f_{\rm S}/f_{\rm EqM}$ in
fig.\,\ref{Fig:Chi_ratio}. Here $f_{\rm S}$ and $f_{\rm EqM}$ are
computed for the same value of $\chi$, being either 0.1, 0.2, 0.4 or
0.6.  Here the bottom curve represents the ratio in the number of
stars for $\chi>0.6$, the two curves give $q_{\rm S/EqM}$ for $\chi >
0.1$.  The ratio $q_{S/EqM} < 1$ for all values of $\chi > 0.1$ and
over the entire time span of the simulations.

The result of two simulations with a mass spectrum (solid and dotted
curves, in fig.\,\ref{Fig:Chi_ratio} show the same behaviour. The
structure of the curves for $\chi > 0.4$ and $\chi > 0.6$ are also
quite interesting. They continue to drop at a constant rate until
model S reaches a second core collapse around $t=60$--70.

The comparison carried out here, though not the main objective of this
paper, is quite interesting from an astrophysical perspective. As the
multi-mass simulation systematically appears to have fewer stars on
irregular orbits. This deficiency can be as small as $q=0.05$ during
core collapse (or there about). We are currently extending this
analysis to larger $N$-body systems on order to study core collapse in
large $N$-body simulations.

\section{Conclusions}\label{Sect:Conclusions}

We present a qualitative diagnostic to analyze irregular behaviour in
particle based gravitational $N$-body simulations.  The method, dubbed
CWaT, is based on continuous wavelet transforms and provides a direct
and instantaneous diagnostic for irregular behaviour.

In our analysis we demonstrate that CWaT is suitable for analyzing
self gravitating $N$-body simulations, as it does not require a long
time series for the analysis and it provides an instantaneous
indicator for the degree of chaos, allowing a quantification of the
number of irregular orbits at any moment in time in a large $N$-body
simulation.

We applied CWaT successfully to analyse gravitational 3-body
interactions. The results of the CWaT method for 3-body systems
compares well with Poincar\'e sections. Here CWaT provides a
qualification of the irregularity of the orbit, whereas for Poincar\'e
sections such an analysis is technically much harder to perform.

Eventually we applied the CWaT method to several N-body systems with
100 particles in which all stars had the same mass, and another set of
simulations in which a mass spectrum was adopted.  The multi mass
system systematically produced fewer stars on irregular orbits,
compared to the equal mass system. In addition, we noticed that during
a deep core collapse the number of irregular orbits dropped even
further, whereas in the relatively low density inter-core collapse
stages the number of irregular orbits tends to increases again.

\section*{Acknowledgements}

We are grateful to Nicolas Faber, Douglas Heggie and Rudy Wijnands for
discussions and comments on the manuscript. We also would like to
thank the anonymous referees for valuable comments and remarks.  This
work was supported by the Netherlands Organization for Scientific
Research (NWO, via grant \#635.000.001), the Royal Netherlands Academy
of Arts and Sciences (KNAW) and the Netherlands Research School for
Astronomy (NOVA).


\appendix
\section*{Appendix: Initial conditions}\label{app:initcond}

The calculations for testing the method are performed with the {\tt
kira} integrator which is part of the {\tt starlab} gravitational
$N$-body environment \cite{SPZ:2001} (see also {\tt
http://www.manybody.org/manybody/starlab.html}).  The initial
conditions used for the 3-body encounters used as examples in
sections\,\ref{subsec:discuss:non_res_pres} and
\ref{subsec:discuss:dem_res_pres} are presented in the tables\,1 and
2, respectively. The numbers in the tables are in standard $N$-body
units \cite{Heggie:1986}.
Exact initial conditions for each of the three encounters studied is
available via {\tt http://modesta.science.uva.nl}.
The CWaT source code is also upon request.

\begin{table*}[!ht]
        \caption{\label{tab:non_res_pres} 
	Initial conditions of the non-resonant preserving 3-body
	interaction. All numbers are in \emph{N}-body units 
        The initial orbital parameters are:
        $m_{1} = m_{2} = m_{3} = 0.5, a = 1, e = 0, b \simeq 1.34, v_{\infty}
\simeq 1.24 v_{c}$ where the critical velocity $v_{c}$ corresponds to a kinetic energy equal to
the binding energy of the binary.
\cite{Heggie:1986}.
}
        \begin{tabular}{c|ccccccc}
                \emph{star}     & mass  & $x$-pos       & $y$-pos       & $z$-pos       & $x$-vel       & $y$-vel       & $z$-vel       \\
                \hline
                $1$             & $0.5$ & $-32.1139669$        & $-19.6019186$        &$2.424416360$         & $-0.170641$ & $0.431148$      & $-0.0226399$          \\
                $2$             & $0.5$ & $-32.6179535$        & $-20.465630$        & $2.424416360$         & $0.665287$ & $-0.117690$       & $-0.0226399$          \\
                $3$             & $0.5$ & $64.73192040$         & $40.067549$         &$-4.84883272$                & $-0.494646$   & $-0.313458$   & $0.0452799$
        \end{tabular}
\end{table*}

\begin{table*}[!ht]
        \caption{\label{tab:dem_res_pres} 
Initial conditions of the democratic resonant preservation 3-body
interaction.  The initial encounter parameters are $m_{1} =
m_{2} = m_{3} = 0.5, a = 1, e = 0, b \simeq 0.28, v_{\infty} \simeq
0.26 v_{c}$.
}

        \begin{tabular}{c|ccccccc}
                \emph{star}     & mass  & $x$-pos       & $y$-pos       & $z$-pos       & $x$-vel       & $y$-vel       & $z$-vel       \\
                \hline
                $1$             & $0.5$ 
	& $11.02969588$         & $-25.8888731$        & $25.3364950$         & $-0.498927$ & $0.206499$      & $-0.0524768$          \\
                $2$             & $0.5$ 
	& $10.74875682$         & $-26.8485987$        & $25.3364950$         & $0.453819$ & $-0.0972697$     & $-0.0524768$          \\
                $3$             & $0.5$ 
	& $-21.7784527$        & $52.7374718$        & $-50.6729901$         & $0.045107$   & $-0.10923$    & $0.104954$
        \end{tabular}
\end{table*}

\end{document}